\documentclass[aps, twocolumn, prl]{revtex4-2}
\usepackage[latin1]{inputenc}
\usepackage{amsfonts, amssymb, braket, mathtools}
\usepackage[caption=false]{subfig}
\usepackage[export]{adjustbox}
\usepackage[colorlinks=true, linkcolor=red , citecolor=blue]{hyperref}
\usepackage[capitalise]{cleveref}
\usepackage{graphicx}
\usepackage[inline, shortlabels]{enumitem}
\usepackage{bbm}
\usepackage{bbold}
\usepackage{float}
\usepackage{array}
\usepackage{chngcntr}
\usepackage[normalem]{ulem}
\usepackage{color}
\usepackage{multirow}
\usepackage{xfp}
\usepackage{siunitx}
\usepackage{booktabs}
\newcommand{\ra}[1]{\renewcommand{\arraystretch}{#1}}

\begin{document}

\title{Boosted Bell-state measurements for photonic quantum computation}

\author{Nico Hauser$^1$, Matthias J. Bayerbach$^1$, Simone E. D'Aurelio$^1$, Raphael Weber$^{2,3}$, Matteo Santandrea$^2$, Shreya P. Kumar$^2$, Ish Dhand$^2$ and Stefanie Barz$^1$}

\email{barz@fmq.uni-stuttgart.de}

\affiliation{$^1$Institute for Functional Matter and Quantum Technologies \& IQST, University of Stuttgart, 70569 Stuttgart, Germany}
\affiliation{$^2$QC Design GmbH, 89081 Ulm, Germany}
\affiliation{$^3$Institute of Theoretical Physics \& IQST, Ulm University, 89081 Ulm, Germany}

\date{\today}

\begin{abstract}
Fault-tolerant fusion-based photonic quantum computing (FBQC) greatly relies on entangling two-photon measurements, called fusions. These fusions can be realized using linear-optical projective Bell-state measurements (BSMs). These linear-optical BSMs are limited to a success probability of 50\%, greatly reducing the performance of FBQC schemes. To improve the performance of FBQC architectures, a boosted BSM scheme taking advantage of ancillary entangled photon pairs and a 4$\times$4 multiport interferometer has been proposed. This scheme allows the success probability to be increased up to 75\%. In this work, we experimentally demonstrate this boosted BSM by using two Sagnac photon-pair sources and a fibre-based 4$\times$4 multiport beam splitter. A boosted BSM success probability of 69.3$\pm$0.3\% has been achieved, exceeding the 50\% limit. Furthermore, based on our BSMs, we calculate photon-loss thresholds for a fusion network using encoded six-ring resource states. We show that with this boosted BSM scheme an individual photon loss probability of 1.4\% can be tolerated, while the non-boosted BSM leads to a photon-loss threshold of 0.45\%.

\end{abstract}

\maketitle

\section{Introduction}
Photonic fusion-based quantum computing (FBQC) is a promising platform for scalable, fault-tolerant quantum computation ~\cite{FBQCBartolucci2023, Raussendorf2003, Rudolph2005, GHZFBQCPhysRevLett.133.050604}. 
The two primitives of FBQC are entangling two-photon measurements, referred to as fusions, and small, pre-entangled resource states of constant size~\cite{FBQCBartolucci2023, Rudolph2015}. 
Quantum gates are implemented by performing fusions between photons of different resource states. The choice of measurement bases in which these fusions are performed dictates the computation.
The success of the computation can then be deduced from the outcomes of the fusions.
Topological fault-tolerance can be achieved by choosing the resource states as well as the fusion network such that the fusion results allow for error detection through parity checks~\cite{FBQCBartolucci2023}. 

The required fusions can be implemented by Bell-state measurements (BSMs), in which the measured photons are projected onto the Bell basis~\cite{loebl2024transforminggraphstatesbell, Rudolph2005}. Since the success probability of these fusions is directly given by the efficiency of the BSM, high BSM efficiencies are desirable for FBQC. Furthermore, efficient BSMs along with encoded fusions (see Figure \ref{fig:fusionsketch}) can be used to improve the tolerance of FBQC schemes to photon loss~\cite{FBQCBartolucci2023}.

BSMs using only linear-optical elements offer scalability due to their low experimental overhead. They are, however, inherently limited to a maximum efficiency of 50\%~\cite{Calsamiglia2001}. In order to overcome this 50\%-bound while staying in the linear-optical regime, schemes using ancillary photons to perform arbitrarily complete BSMs can be realised~\cite{Ewert2014, Grice2011, Bayerbach2023}. These ancilla-boosted BSM schemes are promising candidates for FBQC schemes, as they enable an efficient, scalable implementation of fusions while only relying on additional photons and linear-optical components.

In this work, we demonstrate such a boosted linear-optical BSM scheme based on entangled ancillary photon pairs~\cite{Grice2011}. By using a fibre-based 4$\times$4 multiport interferometer, BSM success probabilities of up to 75\% can be achieved. Using fibre-integrated components circumvents the need for free-space interferometers, thus drastically reducing the experimental complexity in a resource-efficient manner and allowing the integration into existing fibre-based infrastructures. We report experimental BSM efficiencies of 69.3$\pm$0.3\% using this scheme.

\begin{figure}[b]
	\centering
	\includegraphics[width=0.8\linewidth]{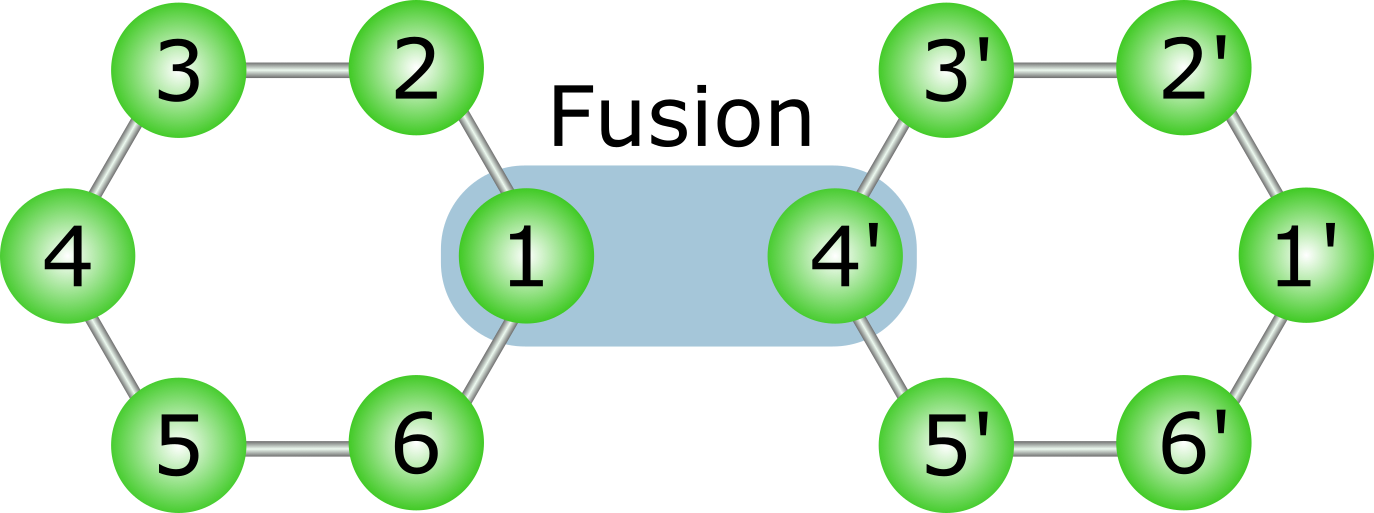}
	\caption{\textbf{Resource states and fusion operation.} The resource states can be represented as a graph with six qubits in a ring configuration. The blue connection represents a fusion operation.}
	\label{fig:sixring}
\end{figure}

\begin{figure*}
	\centering
	\includegraphics[width=0.9\linewidth]{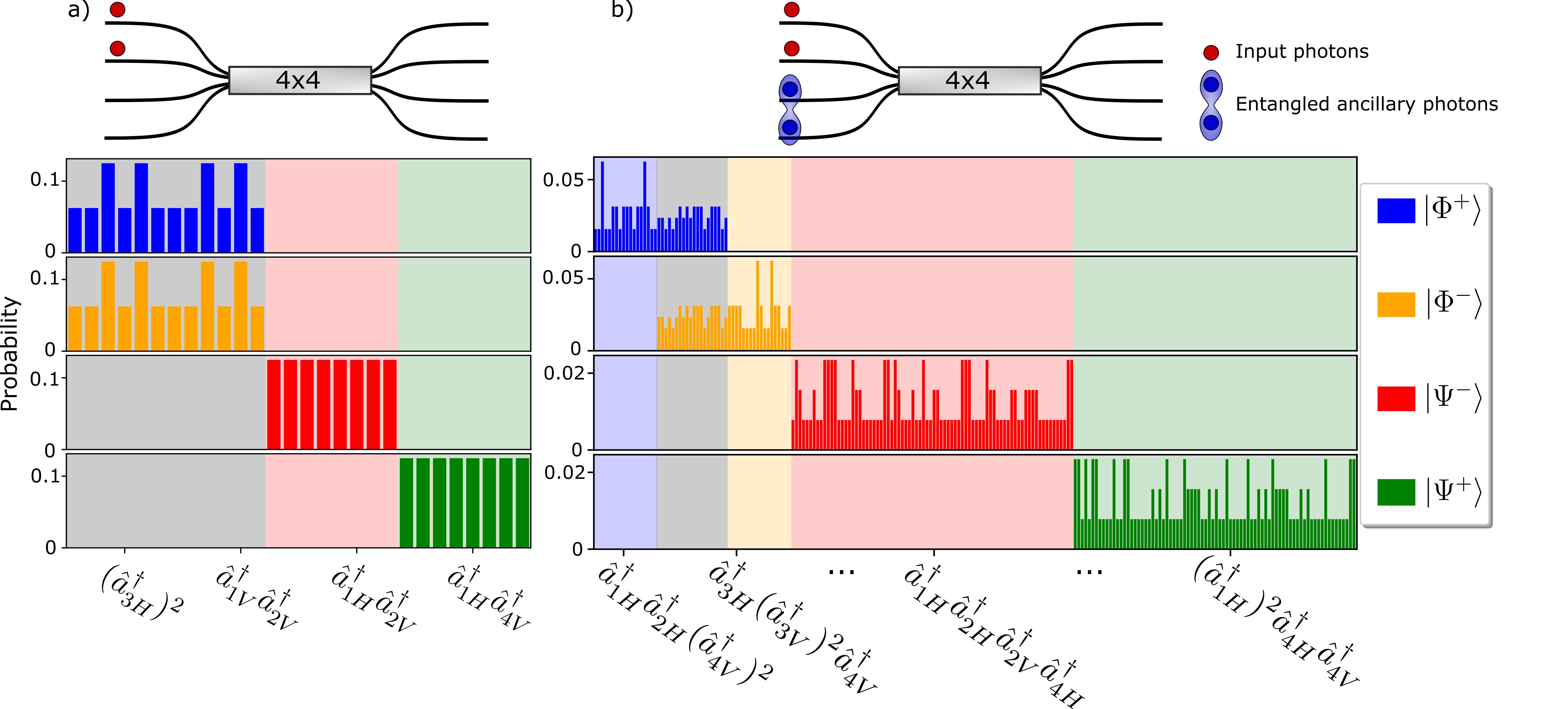}
	\caption{\textbf{Representation of standard and boosted BSM schemes:} \textbf{a)} Schematic and output statistics of the standard BSM scheme using two photons (red circles) as an input for the 4$\times$4 splitter. Each bar corresponds to the probability of measuring a certain detector click pattern. Only the states $\ket{\Psi^+}$ and $\ket{\Psi^-}$ lead to distinct measurement outcomes whilst the states $\ket{\Phi^+}$ and $\ket{\Phi^-}$ cannot be distinguished. The background colours indicate whether a measurement of the corresponding state allows an unambiguous identification of one of the Bell states, with grey denoting ambiguous measurement outcomes. \textbf{b)} Schematic and output statistics of the boosted BSM scheme using an ancillary Bell state (blue circles). The states $\ket{\Psi^+}$ and $\ket{\Psi^-}$ still lead to distinct measurement patterns. Furthermore, the additional interference with the ancillary Bell state gives rise to outcomes that are unique to either $\ket{\Phi^+}$ or $\ket{\Phi^-}$. These unambiguous outcomes allow the identification of these states with a probability of 50\%, consequently leading to a BSM success probability of 75\%. }
	\label{fig:idealstat}
\end{figure*}

Furthermore, we investigate the effects of the boosted BSM on an encoded six-ring fusion network as proposed in~\cite{FBQCBartolucci2023}. Here, the resource state consists of a six-qubit graph state in a ring configuration (see Figure \ref{fig:sixring}), in which each qubit is encoded in four physical photons (see Figure \ref{fig:fusionsketch}). This fusion network has been chosen due to its robustness to imperfect fusions and photon losses~\cite{FBQCBartolucci2023, melkozerov2024analysisopticallossthresholds}.
By simulating photon-loss thresholds for this fusion network, we compare the robustness to photon loss for our boosted BSM to non-boosted BSMs. We show that the boosted BSM allows for an individual photon-loss probability of 1.4\% while the photon-loss threshold for the non-boosted BSM is calculated to be 0.45\%, thus showing a threefold improvement for the boosted BSM.

\section{Theory} 

A linear-optical BSM can be performed by sending two indistinguishable photons to the inputs of a balanced beam splitter and examining the spatial mode as well as the polarisation of the two photons as they exit the beam splitter.

Based on the observed output pattern, the photons are projected into one of the four Bell states $\ket{\Psi^+}$, $\ket{\Psi^-}$, $\ket{\Phi^+}$ and $\ket{\Phi^-}$, where

\begin{equation}
   \ket{\Psi^\pm}  = \frac{1}{\sqrt{2}}\left(\hat{a}_{1H}^\dagger\hat{a}_{2V}^\dagger\pm\hat{a}_{1V}^\dagger\hat{a}_{2H}^\dagger\right)\ket{vac}
\end{equation}

and 

\begin{equation}
   \ket{\Phi^\pm}  = \frac{1}{\sqrt{2}}\left(\hat{a}_{1H}^\dagger\hat{a}_{2H}^\dagger\pm\hat{a}_{1V}^\dagger\hat{a}_{2V}^\dagger\right)\ket{vac},
\end{equation}

with $\hat{a_{i}}^\dagger$ denoting the photonic creation operator with mode number $i$ acting on the vacuum state $\ket{vac}$. 
However, in this configuration, only the Bell states $\ket{\Psi^+}$ and $\ket{\Psi^-}$ produce unique measurement outcomes, whereas the possible outcomes for the Bell states $\ket{\Phi^+}$ and $\ket{\Phi^-}$ are completely identical. This leads to the established standard linear-optical BSM success probability of $p_c=50\%$~\cite{Calsamiglia2001,Michler1996,Bayerbach2023}. The same BSM can be performed by using two inputs of a 4$\times$4 multiport splitter, again leading to a success probability of $p_c=50\%$, as depicted in Figure \hyperref[fig:idealstat]{1 a)}.  \\

For implementing a boosted BSM scheme, an ancillary photon pair in the state

\begin{equation}
	\ket{\Psi_\text{anc}} = \ket{\Phi^+}_{3,4} = \frac{1}{\sqrt{2}}\left(\hat{a}_{3H}^\dagger\hat{a}_{4H}^\dagger+\hat{a}_{3V}^\dagger\hat{a}_{4V}^\dagger\right)\ket{vac}
\end{equation}

is prepared~\cite{Grice2011}.
The input photons and the ancillary photon pair are then sent to the 4$\times$4 multimode splitter. The multimode splitter implements a discrete fourier transform matrix in the form of~\cite{Beige2005}

\begin{equation}
\hat{U}_\text{4x4} = \frac{1}{2} \left(
\begin{array}{cccc}
 1 & 1 & 1 & 1 \\
1 & -1 & -i & i \\
 1 & -1 & i & -i \\
1 & 1 & -1 & -1 \\
\end{array}
\right).
\label{4x4matrix}
\end{equation}

A schematic of the boosted scheme is depicted in Figure \hyperref[fig:idealstat]{1 b)}.
Now, additional terms in the output arise and allow for discrimination between the Bell states $\ket{\Phi^+}$ and $\ket{\Phi^-}$ in 50\% of the cases~\cite{Grice2011}. The expected output statistics for different Bell-state inputs in the 4$\times$4 interferometer are depicted in Figure \hyperref[fig:idealstat]{1 b)}.

In order to get a measure for the quality of the implemented BSM, we use a set of parameters that can be extracted from experimental data, as introduced in~\cite{Bayerbach2023}.
We determine the probabilities $p_c$ and $p_f$ as the probability of identifying the correct or an incorrect input Bell state, respectively. By calculating the average over all possible input Bell states, we obtain $p_\text{c,total}$ and $p_\text{f,total}$ with

\begin{equation}
p_{\text{c/f,total}} = \frac{1}{4}\sum_{\ket{\Psi}} p_\text{c/f}(\ket{\Psi}),  \; \text{for} \; \ket{\Psi}\in\{\ket{\Psi^\pm},\ket{\Phi^\pm}\}.
\end{equation}

In the implemented boosted BSM scheme, $p_\text{c,total}$ can reach a maximum of 75\%. The quantities $p_c$ and $p_f$ allow the definition of the measurement discrimination fidelity (MDF) as~\cite{Wein2016}

\begin{equation}
    \text{MDF}=\frac{p_c}{p_c+p_f},
\end{equation}

which denotes the probability that a measurement uniquely and correctly identifies the prepared Bell state. \\
Another quantity used to describe the quality of the performed BSM is the total variation distance $D$ with~\cite{Pan2019}

\begin{equation}
    D= \sum_i\frac{|f_i-q_i|}{2},
\end{equation}

where $q_i$ and $f_i$ denote the theoretical and measured probability of the click pattern $i$, respectively. \\

Using the measured values for $p_\text{c,total}$ for the boosted and non-boosted BSM, FBQC simulations can be performed to determine thresholds for the \textit{fusion erasure probability} $p_\text{er}$. A fusion is erased if at least one photon used in the fusion is lost, leading to a loss of the measurement outcome.  $p_\text{er}$ can be written in terms of individual photon loss probabilities as

\begin{equation}\label{eq:ploss}
    p_\text{er} = 1-(1-p_\text{loss})^{n_{\text{BS}}+n_{\text{anc}}},
\end{equation}

where $n_\text{BS}$ corresponds to the number of photons to be projected into a Bell state and $n_{\text{anc}}$ is the number of ancillary photons. $p_\text{loss}$ corresponds to the probability of losing an individual photon, which is used as the quantity of interest for the comparison of the different BSM schemes. Basis of the simulation is a fusion network consisting of six-ring graph states as depicted in Figure \ref{fig:sixring}.

Details on the simulation methods are outlined in Appendix \hyperref[app:sim]{E}.

\section{Experiment \& Results}
\begin{figure*}
	\centering
	\includegraphics[width=\textwidth]{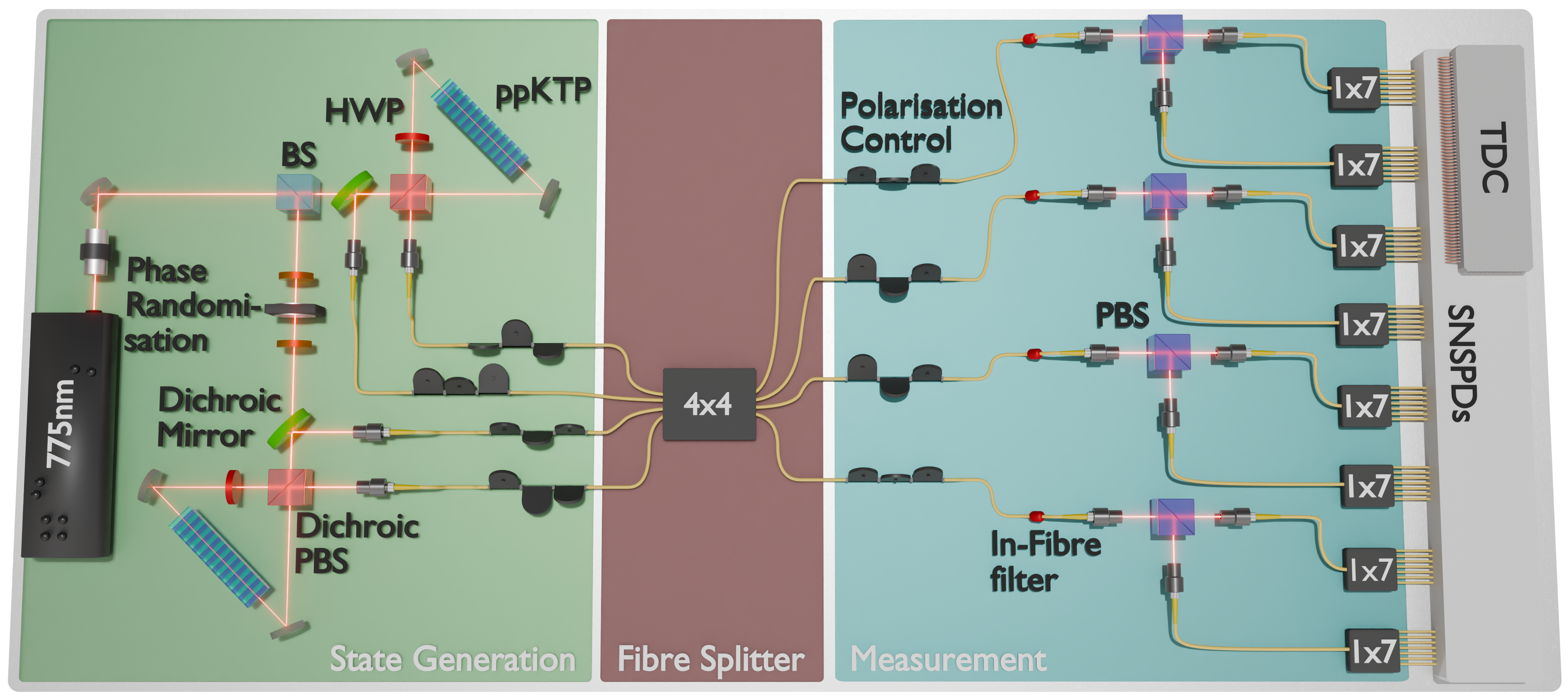}
	\caption{\textbf{Schematic of the experimental setup.} The setup consists of a pulsed \SI{775}{nm} pump laser (\SI{2}{ps} pulse duration, \SI{76}{MHz} repetition rate), which is split using a balanced beam splitter (BS) to pump two Sagnac-type Bell-state sources. The sources are based on spontaneous parametric downconversion (SPDC) in periodically poled potassium titanyl phosphate (ppKTP) and generate polarisation-entangled photon pairs at $\lambda = \SI{1550}{nm}$. For setting the polarisation of the pump laser, half-wave plates (HWP) are used. The generated photonic Bell states are then coupled into single-mode fibres and sent to the fused-fibre 4$\times$4 splitter. The relative phase between the sagnac sources is randomised using two quarter wave plates and a motorized HWP. To ensure that the polarisation of the input photons is maintained, fibre polarisation paddles are used. In the measurement stage, the photons are sorted by polarisation using polarizing beam splitters (PBS) where each output mode is connected to a 1x7 splitter with seven superconducting nanowire single-photon detectors (SNSPD) to allow for probabilistic photon number resolution. For the implementation of this detection scheme, a total of 56 SNSPDs and time-to-digital converters (TDC) are needed.}
	\label{fig:ExperimentSetup}
\end{figure*}

The full experimental setup including two sources of photonic Bell states, the fused-fibre 4$\times$4 multiport splitter and the detection stage is shown in Figure \ref{fig:ExperimentSetup}. \\
In order to analyse the output statistics of the 4$\times$4 fibre splitter, the photons are sent through polarizing beam splitters (PBS). Each output is then probabilistically split up using a 1x7 splitter and sent to superconducting nanowire single-photon detectors (SNSPDs) with an average efficiency of about 80.0\% to achieve pseudo photon-number resolution (PNR).  \\
The statistics for the implemented BSM are measured by recording coincidence events at the detectors. For the improved scheme, a background correction has been performed to account for the probabilistic nature of the Bell-state sources (see Appendix \hyperref[app:background]{B})).

\begin{figure}[b]
	\centering
	\includegraphics[width=\linewidth]{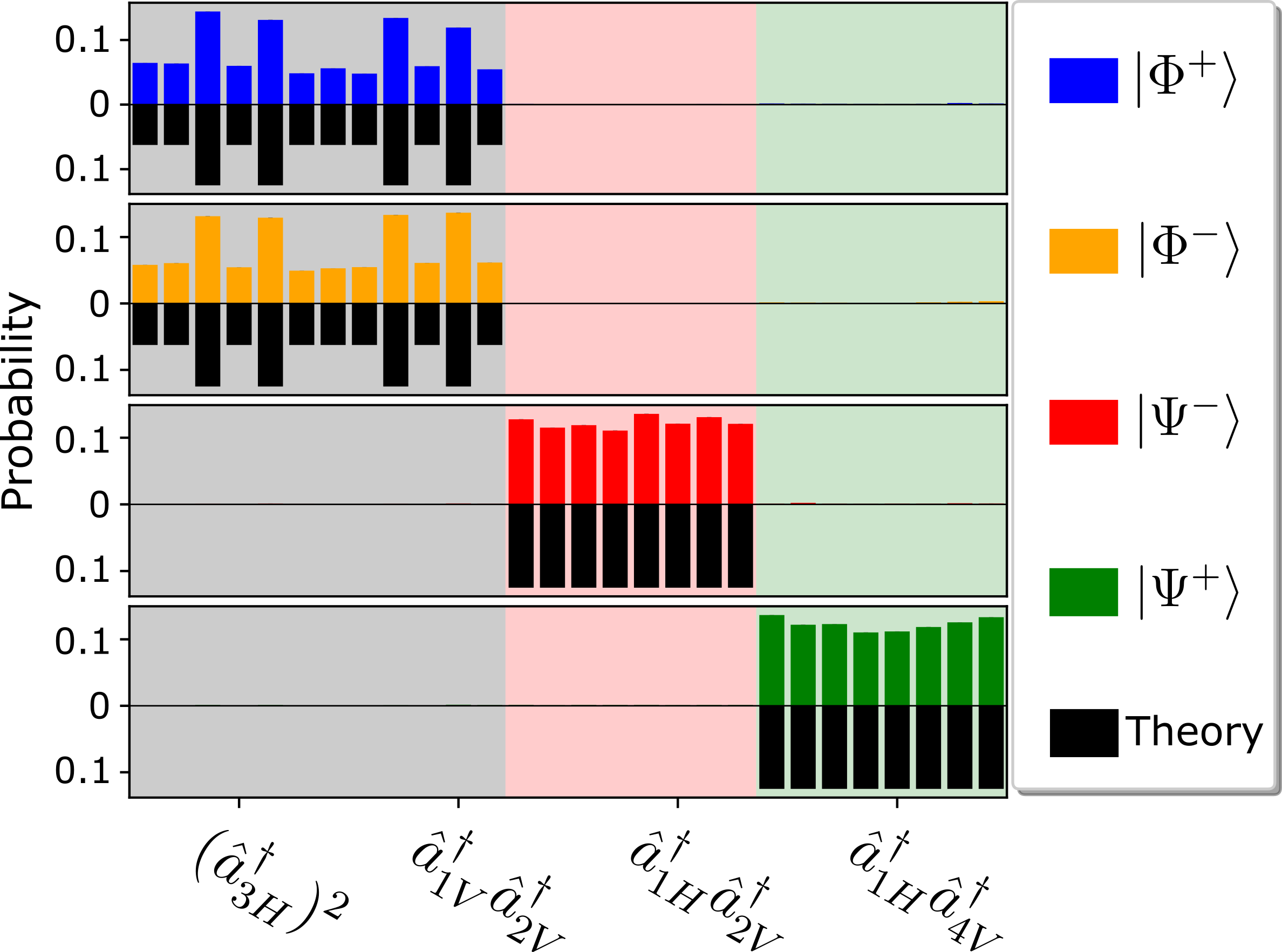}
	\caption{\textbf{Measured probability distribution of two-photon events for the standard BSM scheme.} Only the $\ket{\Psi^\pm}$ states show unambiguous click patterns, while the $\ket{\Phi^\pm}$ states perfectly overlap. Only states that correspond to at least one of the Bell states are depicted. Here, an average success probability of $p_\text{c,total} = (49.05\pm0.02)$\% is achieved. The error on the individual probabilities is smaller than 0.001.}
	\label{fig:standard}
\end{figure}

\begin{figure*}
	\centering
	\includegraphics[width=\textwidth]{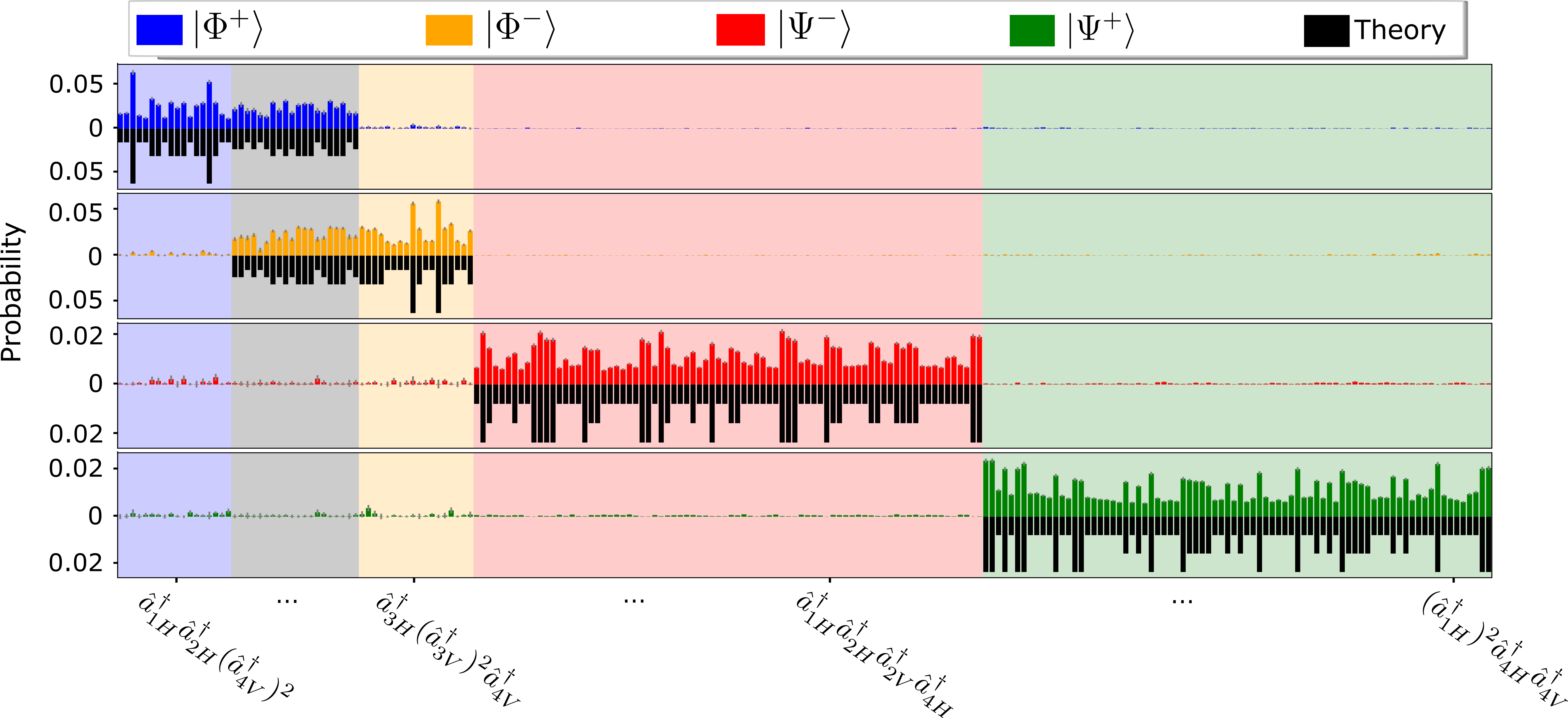}
	\caption{\textbf{Measured four-photon events for the improved Bell-state measurement scheme.} Each bar correspond to the probability of measuring a specific detector click pattern. The states $\ket{\Psi^+}$ and $\ket{\Psi^-}$ can be correctly identified with a probability of 92.4$\pm$0.5\% and 92.7$\pm$0.5\%, respectively. The improved scheme leads to measurement results that are exclusive to either $\ket{\Phi^+}$ or $\ket{\Phi^-}$. These states can be correctly identified with a probability of 46.1$\pm$0.8\% and 46.2$\pm$0.6\%, respectively.}
	\label{fig:correctedresults}
\end{figure*}

\setlength{\tabcolsep}{7.5pt}

\begin{table*}\centering
\ra{1.3}
\begin{tabular}{@{}cccccccc@{}}\toprule
 & \multicolumn{3}{c}{Standard scheme} & \phantom{abc}& \multicolumn{3}{c}{Boosted scheme}\\
\cmidrule{1-8}
Input & \multicolumn{1}{c}{$p_c$} & \multicolumn{1}{c}{MDF} & \multicolumn{1}{c}{$D$} && \multicolumn{1}{c}{$p_c$} & \multicolumn{1}{c}{MDF} & \multicolumn{1}{c}{$D$} \\
$\ket{\Psi^+}$ & 0.9812$\pm$0.0002   &  0.9901$\pm$0.0002 & 0.0393$\pm$0.0001 && 0.924$\pm$0.005 &  0.932$\pm$0.005 & 0.105$\pm$0.004\\

$\ket{\Psi^-}$ & 0.9809$\pm$0.0003 & 0.9889$\pm$0.0003 & 0.0384$\pm$0.0001 && 0.927$\pm$0.005 & 0.938$\pm$0.005 & 0.115$\pm$0.005 \\

$\ket{\Phi^+}$ & N/D & N/D & 0.0541$\pm$0.0001 && 0.461$\pm$0.008 & 0.85$\pm$0.01 & 0.102$\pm$0.007 \\
$\ket{\Phi^-}$ & N/D & N/D & 0.0471$\pm$0.0001 && 0.462$\pm$0.006 & 0.84$\pm$0.01 & 0.105$\pm$0.006 \\
\cmidrule{1-8}
Average &   0.4905$\pm$0.0002 & 0.9895$\pm$0.0003 & 0.0447$\pm$0.0001 && 0.693$\pm$0.003 & 0.890$\pm$0.004 & 0.107$\pm$0.003 \\
\bottomrule
\end{tabular}
\caption{\textbf{Measured parameters of the BSM schemes.} The table shows the measured probabilities $p_c$ of correctly identifying the input Bell state, measurement discrimation fidelity (MDF) and distance $D$ for different input Bell states using the standard and the improved scheme.}
    \label{tab:improved}
\end{table*}

In order to compare the boosted BSM scheme to the standard scheme, we first perform a standard BSM by blocking the ancillary Bell-state source, routing only photons from a single Bell state to the 4$\times$4 splitter. Using this scheme, the states $\ket{\Psi^+}$ and $\ket{\Psi^-}$
can be identified with probability $p_c$ of (98.12$\pm$0.03)\% and (98.09$\pm$0.1)\% and MDF of (98.32$\pm$0.03)\% and (97.0$\pm$0.1)\%, respectively (see Figure \ref{fig:standard} and Table \ref{tab:improved}). As the measured click patterns perfectly overlap between the states $\ket{\Phi^+}$ and $\ket{\Phi^-}$ it is not possible to distinguish them in the standard scheme, hence $p_c$ is not defined. The average distance $D$ is calculated to be $D_\text{avg} = 0.0447\pm0.0001$, indicating a good overlap between experimental data and theory. Note that in Figure \ref{fig:standard}, only measurement patterns corresponding to at least one of the Bell states are shown, while the measurement patterns not corresponding to any of the Bell states are discarded as a failed measurement. These events contribute to 1.37\% of all outcomes.

In order to experimentally verify the boosting of the BSM, the improved scheme with the ancillary Bell state is performed. 
Due to interference with the ancillary Bell state, additional measurement patterns arise. These allow the identification of $\ket{\Phi^+}$ with a probability of (46.1$\pm$0.8)\% and $\ket{\Phi^-}$ with a probability of (46.2$\pm$0.6)\%, leading to an overall BSM-efficiency of $p_\text{c,total} = (69.3\pm0.3)\%$ (see Figure \ref{fig:correctedresults} and Table \ref{tab:improved}). This a significant improvement over the maximum of 50\% using the standard scheme.
The reduction in $p_c$ for the states $\ket{\Psi^\pm}$ compared to the standard scheme is mostly caused by noise from either of the Bell-state sources generating multiple photon pairs at once.
The number of discarded measurement results that do not correspond to any of the four Bell states amounts to 5.45\% of all outcomes.

We can now investigate the robustness of FBQC architectures to photon loss while comparing the standard BSM to our boosted BSM. For this simulation, we consider a six-ring fusion network with $(2,2)$-Shor encoded six-ring resource states, similar to the one proposed in \cite{FBQCBartolucci2023}. In the $(2,2)$-Shor encoding, each logical qubit is encoded in four physical qubits~\cite{shorcode} (see Figure \ref{fig:fusionsketch}). The simulation methods are explained in Appendix \hyperref[app:sim]{E}. \\
As a benchmark for robustness, we investigate the photon-loss threshold. This is done by simulating logical error rates for different fusion-network sizes, which is given by the number of unit lattice cells of the network in each direction, as discussed in \cite{FBQCBartolucci2023}. Hereby, each unit cell consists of two resource states. These simulations are performed while varying the photon-loss probability $p_\text{loss}$. The threshold for performing FBQC is given by the photon-loss probability at which the logical error rate starts decreasing with increasing network size. By surpassing this threshold, arbitrarily low logical error rates can be achieved by scaling the network size accordingly.

\begin{figure}[t]
	\centering
	\includegraphics[width=0.5\textwidth]{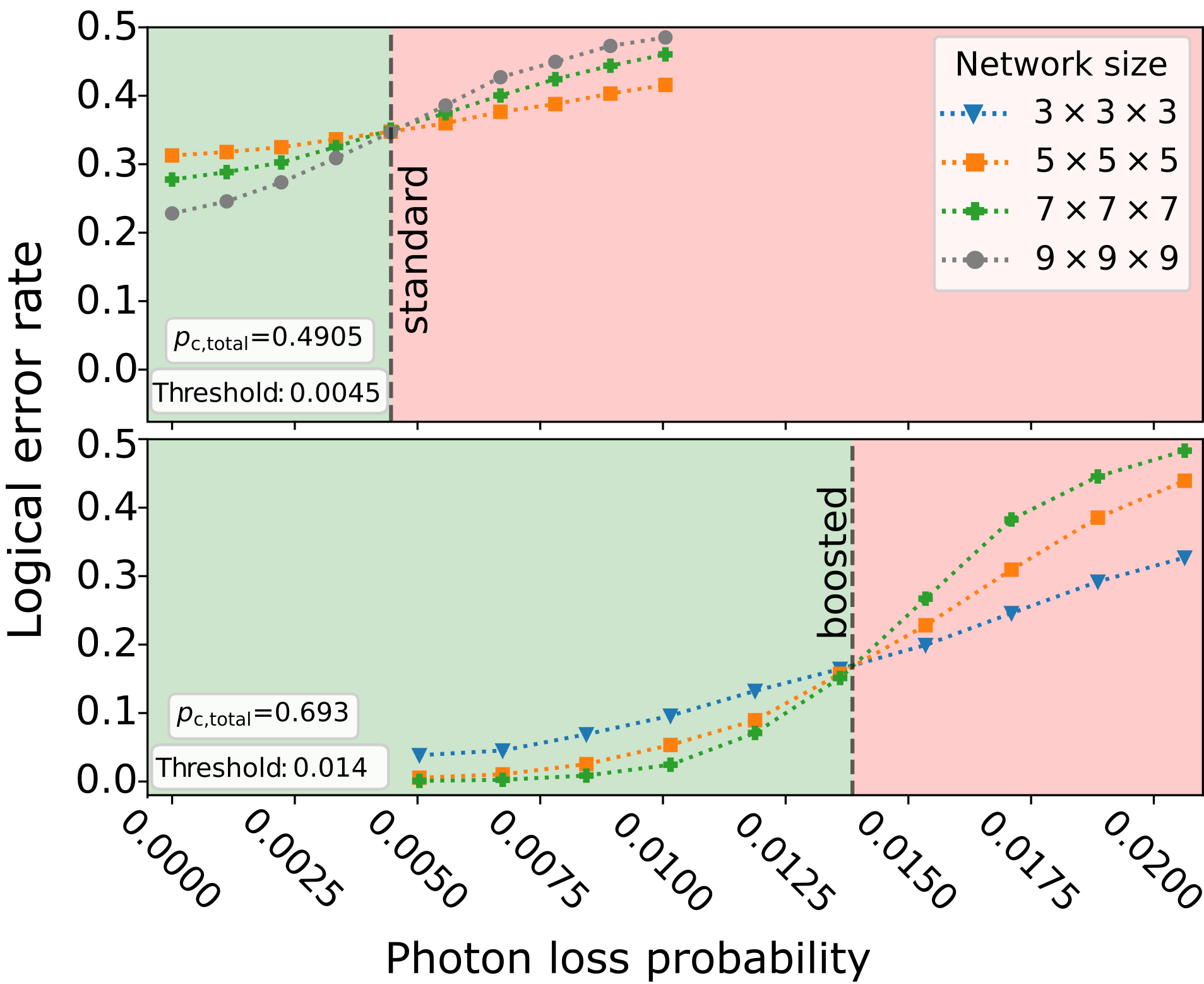}
	\caption{\textbf{Photon-loss thresholds for the standard and boosted BSM.} Logical error rate for different network sizes and photon-loss probabilities for $p_\text{c,total}=0.4905$ (top) and $p_\text{c,total}=0.693$ (bottom). The simulation is performed for different sizes of the fusion network. The intersections mark the threshold at which arbitrarily low logical error rates can be achieved by increasing the size of the fusion network (green background). These photon-loss thresholds are found to be $p_\text{loss}^B=0.014$ for the boosted scheme and $p_\text{loss}=0.0045$ for the standard scheme.}
	\label{fig:Erasure}
\end{figure}

The simulation results for network sizes three, five, and seven for fixed BSM success probabilities of $p_\text{c,total}=69.3\%$ for the boosted scheme and network sizes five, seven, and nine for BSM success probabilities $p_\text{c,total}=49.05\%$ for the non-boosted BSM can be seen in Figure \ref{fig:Erasure}. Here the photon-loss thresholds are determined to be $p_\text{loss}^B = 1.4\%$ for the boosted BSM and $p_\text{loss}=0.45\%$ for the standard BSM. Therefore, using the boosted BSM scheme, the robustness to photon loss of FBQC schemes is significantly improved by more than a factor of three.

\section{Conclusions}
We present the implementation of a boosted BSM scheme using entangled ancillary photon states. We significantly surpass the usual 50\% BSM-efficiency limit for linear-optical schemes and find improved photon loss thresholds for FBQC schemes. 
With an average success probability of (69.3$\pm$0.3)\%, we reach the highest ancilla-assisted linear-optical BSM success probability that has been demonstrated to our knowledge. We show that using the boosted BSM, fusion-based photonic quantum computation could be performed with an individual photon-loss probability as high as $p_\text{loss}^B = 1.4\%$, compared to $p_\text{loss} = 0.45\%$ with the standard BSM scheme. Thus, a significant increase in the loss robustness for fusion-based architectures can be achieved.
Using fibre-integrated multiport splitters, we eliminate the need for complex free-space interferometers and thus greatly reduce the experimental complexity as well as required components compared to bulk-optical implementations of similar schemes.  \\
Our results show the benefit of using pre-entangled ancillary photon pairs as a resource to increase the success probability of Bell-state measurements, surpassing schemes in which non-entangled ancillary photons are used~\cite{Bayerbach2023,Ewert2014}. Due to the significance of efficient Bell-state measurements in photonic quantum computation, this work marks a milestone towards the viability of such architectures in real-world applications~\cite{Li2015,Rudolph2007,Pant2019,Gottesman1999,DANOS200773}.

\section{Acknowledgements}
We thank Shreya Kumar and Daniel Bhatti for helpful discussions and comments.\\
We would like to acknowledge support from the Carl Zeiss Foundation, the Centre for Integrated Quantum Science and Technology (IQST), the Deutsche
Forschungsgemeinschaft (DFG, German Research Foundation) -
431314977/GRK2642
 and the Federal Ministry of Education and Research (BMBF, projects SiSiQ and PhotonQ).

\appendix

\section{Appendix}
\renewcommand{\thefigure}{A\arabic{figure}}
\setcounter{figure}{0}

\subsection{A) Photon indistinguishability}
In order to verify that the generated photons are indistinguishable in their spectral and temporal degree of freedom, Hong-Ou-Mandel experiments~\cite{HOM1987} have been performed between two photons being emitted from the same source as well as heralded single photons being emitted from different sources. Here, we achieved interference visibilities of $V_\text{HOM1} = (99.2\pm 0.3)\%$ and $V_\text{HOM2} = (99.0\pm0.4)\%$ for photon pairs emitted from the first and second source, respectively, while pumping the sources with \SI{50}{mW}. For heralded single photons from each of the sources, we achieved an interference visibility of $V_\text{HOM12}=(94.0\pm 0.6)\%$, indicating the generation of pairwise indistinguishable photons from both sources. Here, the two-source visibility is slightly lower compared to the single-source visibility due to additional noise from higher-order emissions as well as residual spectral correlations. \\

\subsection{B) Background correction}\label{app:background}
The Bell-state sources used in this work are based on SPDC and as such generate two-mode squeezed vacuum states~\cite{lvovsjy2015}. Consequently, higher-order emissions may occur, making a background correction necessary.
The background is corrected for by measuring and then subtracting the output statistics while blocking either Source.
Furthermore, the relative phase between the two sources is randomised to circumvent induced coherence effects~\cite{Mandel1991}.
Employing a background correction can lead to small negative values for measurement outcomes close to zero. In these cases, the absolute values are taken, leading to a slight underestimation of the reported BSM efficiencies.
The need for background correction can be eliminated by using Bell-state sources based on deterministic single photon sources~\cite{pennacchietti2023} or heralded Bell-state generation schemes~\cite{Dhara2022} in future implementations.\\

\subsection{C) Pseudo photon-number resolution}
To achieve pseudo photon-number resolution, we probabilistically split each output mode into 7 modes. \\
The outputs are split using fibre-based balanced 1x7 splitters and connecting each of its outputs to an SNSPD. The probability of $n$ input photons being routed to different detectors in the case of $k$ outputs can be calculated as

\begin{equation}
    P(n,k) = \frac{k!}{(k-n)!k^n}.
\end{equation}

Since this scheme is inherently probabilistic, the measured probabilities of events in which multiple photons are sent into an 1x7 splitter are corrected by a factor of $1/P_\text{PPNR}$ with

\begin{equation}
    P_\text{PPNR}= \prod_{i=1}^7 P(n_i,k),
\end{equation}

where $n_i$ is the number of photons incident on the 1x7 splitter in mode $i$.

\subsection{D) Uncertainty calculations}
The data presented in this work is generated by repeatedly measuring coincidence events in 120 second intervals. In order to achieve statistical significance, these measurements are repeated at least 500 times. \\
In order to calculate the uncertainty for the number $N_i$ of events in which click pattern $i$ is measured, the standard deviation

\begin{equation}
    \sigma_{N_i} = \sqrt{\frac{1}{n}\sum_{k=1}^n(N_{i,k}-\Bar{N_i})^2}
\end{equation}

is calculated and used as the uncertainty $\Delta N_i$. In order to determine the uncertainty of all values calculated from $N_i$, Gaussian error propagation is used.

\subsection{E) Simulation methods}\label{app:sim}

We performed FBQC simulations using Plaquette, an end-to-end fault-tolerance simulation software capable of modeling hardware imperfections in photonic and matter-based hardware platforms.

The six-ring fusion network (FN) simulated in this paper corresponds to the one discussed in \cite{FBQCBartolucci2023}, where we assume periodic boundary conditions. 
The arrangement of the resource states in this network can be described by two distinct lattices: the primal and the dual lattice. These lattices are useful to detect the different error mechanisms. For example, an X (or Z) error that affects the primal lattice will not affect the dual lattice and vice versa, whereas a Y Pauli error will affect both lattices at the same time. 
In this way, the checks of the six-ring FN discussed in \cite{FBQCBartolucci2023} can be separated into two decoupled graphs, the primal and dual decoding graph, which are automatically extracted by Plaquette. In each decoding graph, each node corresponds to a check of the FN, and each edge connecting two nodes corresponds to the fusion measurement outcome whose error will flip the parity of the two associated nodes. During the simulation, for each fusion, we decide which measurement outcome is lost upon fusion failure, so that its related edge in the decoding graph will be erased (i.e. randomized) In the case of fusion erasure, both edges (one in the primal and the other in the dual graph) are erased. Therefore, in the presence of both fusion failure and erasure, a subset of edges in both the primal and the dual decoding graph are erased, thus changing the parity of the checks connected to them. The checks with inverted parity constitute the syndrome, which provides information on where errors might have occurred. This syndrome is passed to the FusionBlossom~\cite{wu2023fusionblossomfastmwpm} decoder available in Plaquette, which provides a possible correction. To assess whether a logical error has occurred, the parity of the correlation surface in the absence of errors and after the error correction procedure is compared: if the two parities are different, then a logical error has occurred.

\begin{figure}
	\centering
	\includegraphics[width=0.4\textwidth]{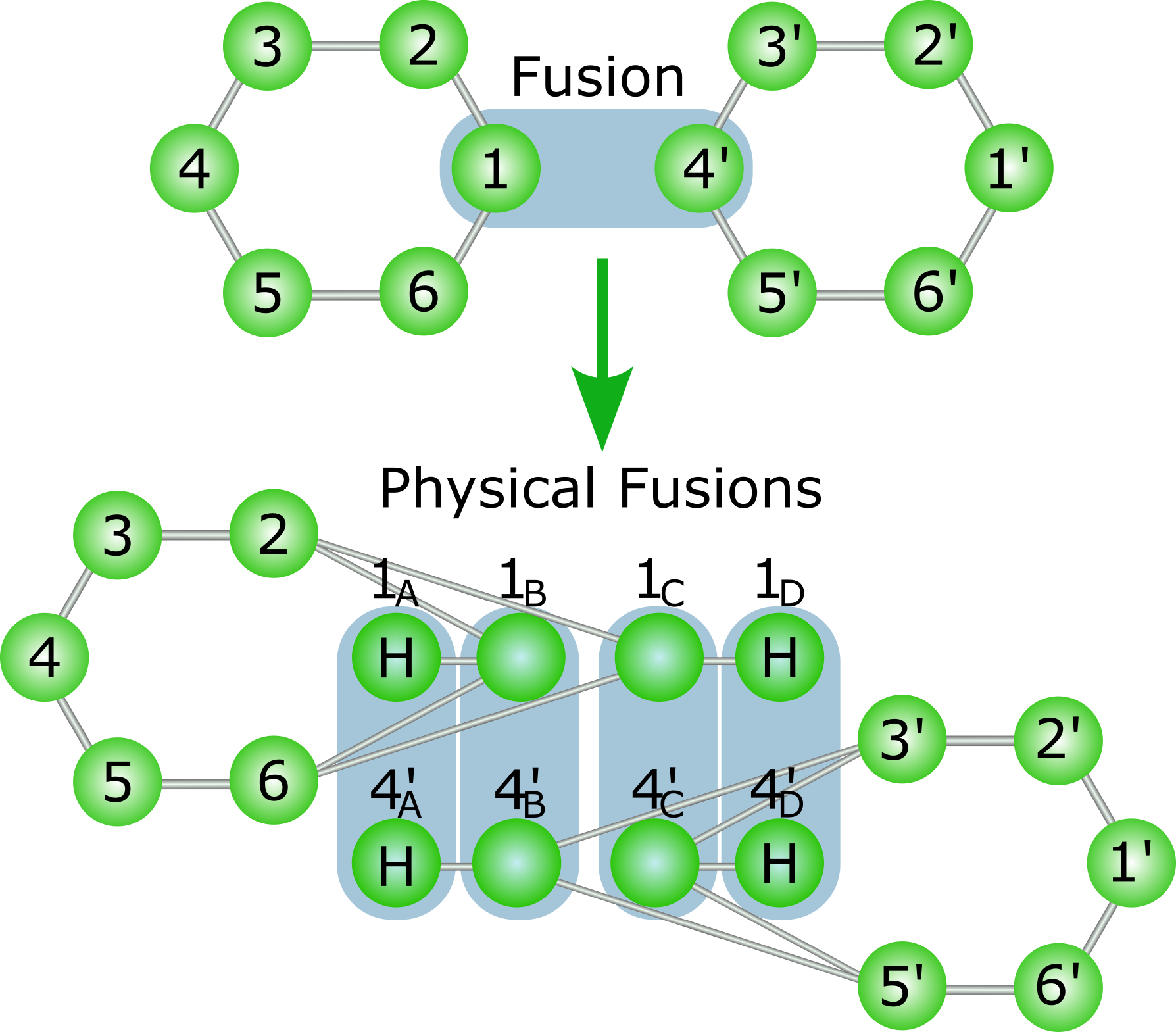}
	\caption{\textbf{Encoded fusion (blue connection) between qubits (green circles) of two six-ring graph states.} (Figure adapted from ~\cite{FBQCBartolucci2023}). Using the $(2,2)$-Shor encoding, each logical qubit is encoded in four entangled physical qubits, denoted with index A to D. All physical fusions measure the parity of the stabilizer generators $XX$ and $ZZ$ on the physical qubits $1_\text{i}$ and $4_\text{i}^\prime$. Upon fusion failure, we lose the parity of $ZZ$ in fusions A and C and the parity of $XX$ in fusions B and D, guaranteeing a fusion failure probability of the encoded fusion of $p_\text{fail, enc} = p_\text{fail, phys}^2$, where $p_\text{fail, phys}$ denotes the failure probability of the physical fusion. If there are no other sources of error (apart from fusion failure), we always lose the $ZZ$ measurement outcome of the encoded fusion if it fails.}
	\label{fig:fusionsketch}
\end{figure}

For the simulations related to the $(2,2)$-Shor encoded six-ring fusion network, we mapped fusion failure and erasure probabilities on physical fusions to fusion failure and erasure probabilities on the encoded fusions. This mapping was performed in similar fashion to the mapping presented in \cite{FBQCBartolucci2023}. The main difference in our simulation is that we consider the specific pattern of fusion measurements that maximizes the fusion success probability of the encoded fusion for the given type of encoding chosen. An implementation of such an encoded fusion from physical fusions is shown in Figure \ref{fig:fusionsketch}.

For the specific threshold plots that are depicted in Figure \ref{fig:Erasure}, we fix the fusion success probabilities $p_\text{c,total}$ and scan over erasure probabilities. In more detail, we choose a set of physical erasure probabilities, which we scan over. These probabilities along with the fusion success probabilities are then translated into effective erasure probabilities on encoded fusions. Logical error rates are computed repeatedly for these effective erasure rates for different sizes of FBQC lattices (comprising $n \times n \times n$ unit cells for $n$ in $3, 5, 7$). In the end, the obtained logical error rates are plotted against the different photon-loss probabilities and a threshold is obtained in the presence of fusion failure. Here, we assume uniform loss acting on all photons involved in the fusion, namely two in the case of unboosted BSMs and four in the case of boosted BSMs

Based on this analysis, the loss threshold for the $(2,2)$-Shor encoded six-ring fusion network with a fusion success probability of $69.3\%$ is $p_\text{loss}^B = 1.4\%$, and the loss threshold for the same fusion network with a fusion success probabilities of $49.05\%$ is $p_\text{loss} = 0.45\%$.

\bibliography{GriceBib}

\end{document}